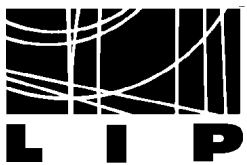

# HIGH RESOLUTION RPC'S FOR LARGE TOF SYSTEMS

P. Fonte[1,3,#], R. Ferreira Marques[2,3], J. Pinhão[3], N. Carolino[3], A. Policarpo[2,3]

1 – CERN, European Laboratory for Particle Physics, CH-1211, Geneve 23, Switzerland.
2 – Departamento de Física da Universidade de Coimbra, 3004-516 Coimbra, Portugal.
3 – LIP – Laboratório de Instrumentação e Física Experimental de Partículas, Coimbra, Portugal

## Abstract

Here we report on a particular type of RPC that presents up to 99% efficiency for minimum ionizing particles and a very sharp time resolution, below 50 ps $\sigma$ in the most optimized conditions. Our 9 cm$^2$ cells, made with glass and metal electrodes that form accurately spaced gaps of a few hundred micrometers, are operated at atmospheric pressure in non-flammable gases and can be economically produced in large quantities, opening perspectives for the construction of large area time of flight systems.

# 1. - Introduction

Heavy-ion collision physics at very high energies is emerging in many accelerator centers around the world (RHIC at BNL, SIS at GSI, and LHC-HI at CERN), emphasizing the need for improved particle identification systems. Many among the particles produced in huge numbers in such processes have their momentum spectra within the range covered by Time of Flight (TOF) techniques.

Our team has a longstanding interest in Parallel Plate Chambers [1] and in recent years carried on further research work along complementary lines related to both the in depth understanding of processes occurring in gaseous detectors of parallel plate geometry [2] and the behavior of detectors equipped with resistive electrodes [3].

In the framework of the R&D program for the ALICE TOF system[a] in which we are involved [4], and following progress made by the ALICE-PPC team, we designed, constructed and beam-tested an RPC-like device, which consists of a pair of double-gap Resistive Plate Chambers (RPC) and seems suitable for the construction of large area TOF arrays. Among its main properties, one may stress the small current pulses in case of discharge, the high detection efficiency, the high geometrical efficiency, and, particularly, very good time resolution.

In this paper we describe the structure of such detector and present results from single-cell tests carried out at the CERN's PS T10 beam. An array of 32 cells was also tested recently by the ALICE-TOF project with encouraging results [5].

## 2. - **Description of the detector cells**

RPCs with internal floating electrodes made with phenolic laminate were first reported several years ago [6] and since then actively developed by the same group. We, instead, use accurately spaced gas gaps delimited by rigid glass and metal electrodes.

Our detector cells, schematically represented in Fig.1, are constituted by a pair of identical double-gap chambers, each one having metallic anode and cathode (aluminium) and a central resistive electrode operated at floating potential. The cathodes are connected to negative high voltage and the anodes, kept at ground potential, are connected together and feed the pre-amplifier. The glass electrode, made from a commercially available darkened glass, has a resistivity $\rho \approx 2 \times 10^{12}$ $\Omega$cm. The thickness of the plates was 2 mm for the aluminium ones and 3 mm for the glass.

In order to reduce dark counts and avoid discharges, the edges of the electrodes, particularly those facing the gaps, were all wedged at an angle of aprox. 30 degrees (in the case of the glass) or rounded off (the approximate radius of the aluminum electrode edges was 0.3 mm).

After machining, the aluminium electrodes were slightly polished and their flatness estimated by an expedite capacitive method. A small number of plates was accurately measured, showing that the metal and glass plates have a flatness generally better than

---

[a] The ALICE Experiment at the LHC-HI program plans to install a TOF barrel system with about 100 m$^2$, located at 3,5 m radius and covering the ±1 rapidity range, whose goal is the identification of particles from 1,4 to 2 GeV/c (97% of the emitted particles will have $p_t$<2 GeV/c). Such a TOF system must present good time resolution and fine segmentation (of the order of $10^5$ channels), together with the capability of coping with moderate counting rates, ≈$10^2$ Hz/cm$^2$.

4 μm and 2 μm, respectively, and a typical surface roughness (*Ra*) around 0.6 μm and 0.02 μm, respectively.

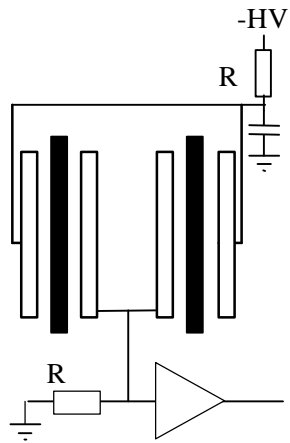

Figure 1 - Schematic view of the detector cell and readout method. The long white rectangles represent the aluminum electrodes and the black ones the resistive glass electrodes.

The different elements are assembled in a plastic box that provides electrical insulation and mechanical rigidity (Fig. 2). This box has holes for the wires that connect each metallic electrode, as well as for the gas circulation and for the insertion of the spacers that define the gas gaps. These spacers are tiny pieces of optical glass fiber of $\phi=300$ μm, that are inserted along the four corners of the chamber. The picture also shows an extra Plexiglas spacer that separates the two double-gap chambers of a cell.

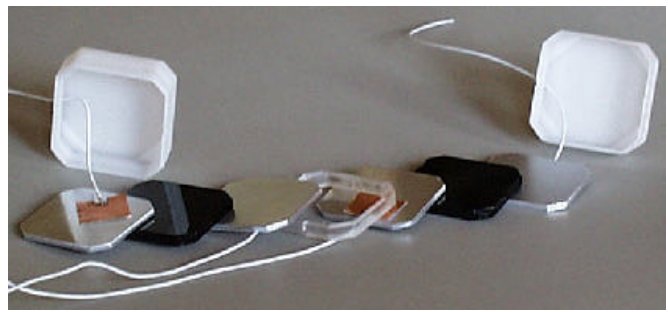

Figure 2 - Relative position of the various elements that compose one cell. The spacers (glass fibers) are the only elements not shown in the picture.

A sample of the glass fiber spacers was also accurately measured, being found that the it's diameter typically lies within ±3 μm of the nominal value.

In the present design, the cells present a thickness corresponding to 5.5% of a radiation length, but this figure is expected to drop by a factor two with the use of thinner glass and aluminium plates.

The simplicity of construction and the easy availability of the materials render this type of cells quite suitable for economic mass production.

The electrodes are octagonally shaped in order to permit a denser packing of the cells (starting from squares with 32 mm long sides, 3 mm were removed from each corner). The assembly is made in a chessboard-like pattern, as it can be seen in Fig. 3. A second set of cells is then installed on top of the first layer, displaced by one pitch unit (3 cm in this case), resulting in a geometrical coverage better than 95% for normal incidence.

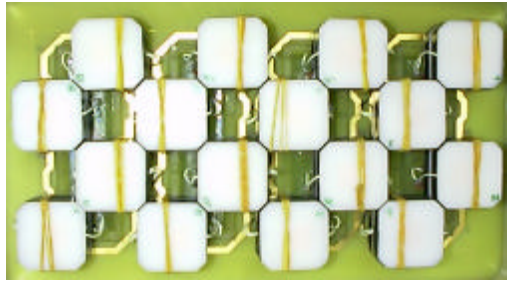

Figure 3 - Array of 16 cells mounted in a chessboard-like pattern over a printed circuit board of the ALICE-TOF project. A second layer, placed on top of this one and displaced by one pitch unit, provides a geometrical coverage better than 95%.

If need arises, a metallic shielding can be placed between adjacent cells or the outer surface of the plastic boxes be coated with a conductive layer in order to reduce inter-cell crosstalk.

## 3. - Experimental setup and data analysis

The detector cells were installed in a gas tight box that was kept under continuous flow of a non-flammable gas mixture consisting of $C_2H_2F_4$+10%$SF_6$+5%isobutane [7]. The tests were carried out at the CERN PS, using a secondary beam (T10) of 7 GeV particles, mainly negative pions, from May to July 1999. The spills were 0.3 s long, spaced by a few seconds. The beam was defocused, illuminating the detector cells uniformly, the data acquisition system being triggered only when the particles hit the detectors in a central area of 1 cm$^2$, defined by a scintillator telescope.

The chamber signals were amplified by a simple common-emitter input stage voltage pre-amplifier that we developed especially for this application, based on the Philips BFG520 transistor. It has a 2 ns rise-time, an overall gain of 0.6 V/pC and an equivalent noise level of 0.2 fC. A commercial chip (Philips NE5204) provides a further factor 10 amplification.

The amplifier was followed by a fixed threshold discriminator connected to a TDC with a 50 ps bin width. The discriminating threshold was optimized for time resolution at a counting rate during the spills of 1 kHz/cm$^2$, being 20 fC for the chamber with 0.3 mm gaps and 3 fC for the chamber with 0.2 mm gaps. This contrast is probably due to different noise levels of the two electronic channels used, rather than to any difference in the chambers behavior.

The output of the amplifier was also sent to a charge integrating ADC that measured the fast signal charge. This information was used offline to correct for the finite rise time of the amplifier (slewing correction).

The contribution of the electronics chain to the timing jitter was measured by injecting in the pre-amplifier input some electronically generated pulses containing a random amount of charge, ranging from 0 to 3 pC. The electronics jitter thus measured was around 25 ps after slewing corrections.

Two configurations of counters were used: a minimum of two chambers under test and a very high time resolution scintillator-based counter (T0 counter) or a single chamber under test and two scintilator-based counters (T1, T2 counters). The first configuration was used in to test the chambers referred below as *A* and *B*, and the second configuration to test the chamber *C*.

The individual time resolutions of these counters were estimated by applying the following method to the set of chambers plus the T-counters:

a) For a total number of *N* detectors the set of $(N^2-N)/2$ possible time differences between detectors *i* and *j* was experimentally measured. The corresponding variances ($V_{ij}$) were determined by fitting the distributions with a gaussian curve within $\pm 2\sigma$.

b) The equations $\{V_{ij} = V_i + V_j\}$ were solved for the individual variances $V_i$. If *N>3* the system is over-determined and the solution should be taken in the least-squares sense. The case *N=2*, corresponding to the common START-STOP method, is under-determined and to extract the individual detector resolutions requires further assumptions (e.g., $V_i = V_j$).

The method proved to be quite robust, yielding a resolution of 25 to 30 ps for the T0 counter and from 45 to 50 ps for the T1 and T2 counters, quite independently of the variations introduced in the chamber's operating conditions.

## 4. - Results

We studied the behavior of chambers as a function of rate and applied voltage, for gap sizes of either 300 or 200 μm, hereafter called chambers A and B, respectively. A second 0.3 mm chamber (C), made using improved polishing methods and thinner (1.5 mm) glass plates was also tested.

A typical fast charge spectrum of chamber C for minimum ionizing particles is shown in Fig. 4. It is clear that such pulse height spectrum allows for a very high detection efficiency (99 %) when the threshold is placed as low as 0.02 pC.

Taking in consideration that the gas gaps are very thin and that the final charge contained in the avalanches depends exponentially on the position of the ionization clusters that initiated them, it seems unlikely that the observed pulse height spectrum can be explained without invocation of some unusual process. Investigations are under way to clarify this subject.

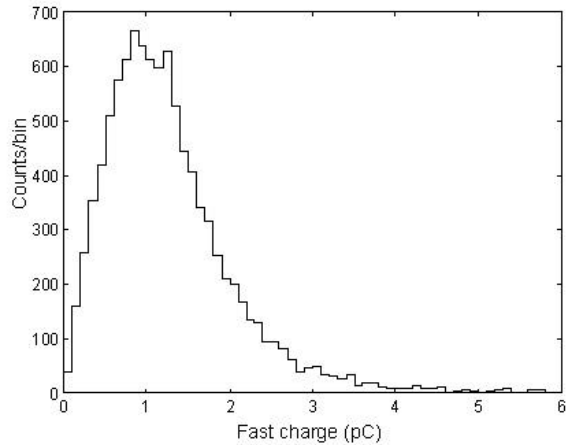

Figure 4 - Fast charge spectrum (electron component) for MIPs in chamber C (similar shapes were observed for chambers A and B), corresponding to a detection efficiency of 99 %. The counting rate during the spills was 300 Hz/cm$^2$.

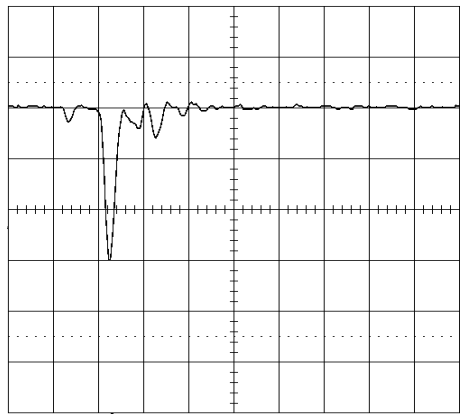

Figure 5 - Typical streamer current pulse for chamber A directly observed on a 50 Ω resistor. The horizontal scale corresponds to 20 ns/division and the vertical scale to 1 mA/division, representing each square a charge of 20 pC. A clear precursor avalanche is visible 15 ns before the main pulse which corresponds to the development and quenching of the streamer [1,2].

A small fraction of the particles triggered the formation of streamers that develop a moderate amount of charge, as shown in Fig. 5.

A typical time difference distribution for chamber C is shown in Fig. 6. As stated above, the distribution was fitted within ± 2σ (solid line) and the corresponding curve extrapolated to the 3σ level (the y axis was adjusted such that the bottom of the picture corresponds to ± 3σ). The distribution is symmetric down to the 2.5σ level and a slight asymmetry can be seen close to the 3σ level. At the present stage it is not clear whether this asymmetry is an unavoidable feature of the detector or an artifact generated by the measuring system (beam, T-counters and electronics).

In Fig. 7 we present time resolution and detection efficiency data as a function of the applied voltage at a counting rate (during the spill) of 1 kHz/cm$^2$. The detection efficiency was defined as the ratio between the number of triggers for which a valid timing signal from the chambers was recorded and the total number of beam triggers recorded. The time resolution was determined by the method described above.

If we define the working plateau as the voltage range for which the time resolution is below 80 ps and the detection efficiency above 95 %, then chambers A, B and C show plateaus of 1.7 kV, 1.0 kV and 1.5 kV, respectively.

As for the time resolution, the best value measured was 44 ps $\sigma$ for chamber C (see Figs. 6 and 7).

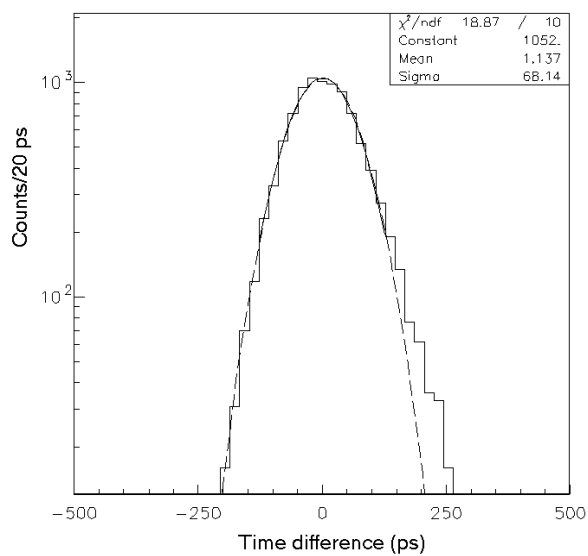

Figure 6 - Distribution of the time difference between chamber C and the T1 counter, showing a width of 68 ps $\sigma$. After quadratic subtraction of the estimated resolution of the T1 counter one gets a resolution of $\sqrt{67^2-49^2}=47$ ps $\sigma$ for the RPC. The counting rate during the spill was 500 Hz/cm$^2$ and the applied voltage 5800 V.

It has been experimentally determined that, for gas gaps of 0.3 mm or larger, the timing jitter in parallel-plate detectors varies almost linearly with the width of the gaps [4]. The present data suggests that below 0.3 mm this trend may not be followed. In addition, chambers with 0.1 mm gaps were also tested, but have shown no improvement in time resolution. Since a mechanically improved version (C) of the standard 0.3 mm gap chamber (A) has shown a sharper time resolution this may be ascribed to the electrode planarity deviations that, in relative terms, become more important for the shorter gaps. If this is the case, then chambers built with tighter mechanical tolerances and narrower gaps may be able to reach event better timing accuracy.

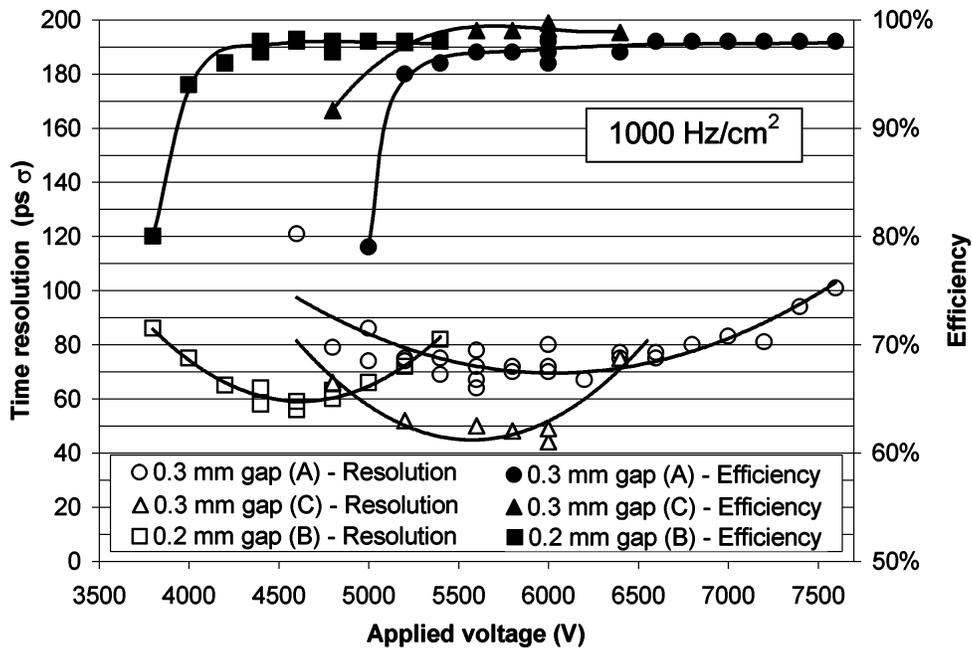

Figure 7 - Efficiency and time resolution as a function of the applied voltage. Arbitrarily shaped curves (splines) were fitted to the efficiency data and parabolic curves to the resolution data.

In Fig. 8 we present time resolution and detection efficiency data as a function of the average counting rate during the spills. The degradation observed for higher counting rates, is probably due to variable voltage drops across the resistive plate that affect the local gain in a time-dependent manner. Indeed the pulse height spectrum changes progressively with increasing rate from the peaked shape shown in Fig.6 to an exponential distribution. In this case, and since detectors A and B are identical, except for the gap width, the superior rate capability of chamber B may be attributed to the lower discriminating threshold used in this chamber. In principle, the use of thinner resistive plates (as in chamber C) and lower resistivity materials will increase the rate capability of these chambers.

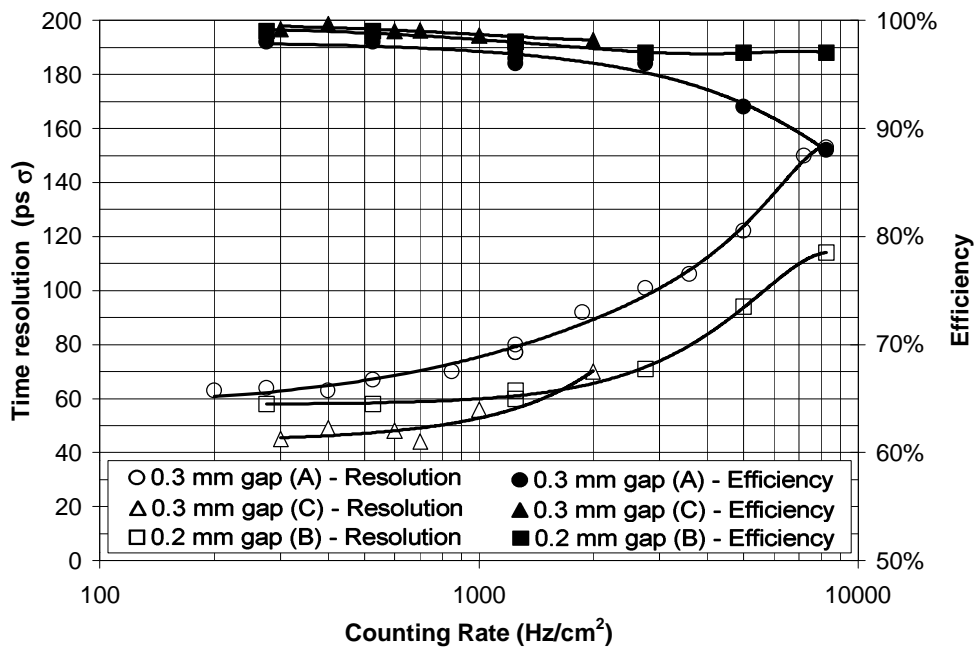

Figure 8 - Efficiency and time resolution as a function of the average counting rate during spills. Arbitrarily shaped (spline) curves were fitted to the data points.

## 5. – Conclusions

We built and tested Resistive Plate Chambers with an active area of 9 cm$^2$ made of rigid, stable in time and accurately spaced metal and glass electrodes, forming a pair of double gap chambers with gaps of 0.3 or 0.2 mm.

The best time resolution measured was 44 ps σ at a detection efficiency of 99%. The detectors kept resolutions below 80 ps σ and detection efficiencies above 95% along an operating voltage range of more than 1 kV and for average counting rates during beam spills up to 1 kHz/cm$^2$. The use of thinner resistive plates and lower resistivity materials should increase the rate capability of these chambers [3].

The optimum discriminating threshold for the 0.3 mm gaps chamber was 20 fC, well above the 0.2 fC electronics noise level and much below the average fast pulse charges (around 1 pC), providing a large safety margin against system-generated noise or crosstalk from the neighboring cells.

The main features of these detectors, simple and mechanically stable construction, very good time resolution and detection efficiency, large plateau and high discriminating threshold, seem to render this approach suitable for the mass-scale production of cells for large area TOF arrays.

When compared with scintillator-based TOF solutions, the present approach has the advantages of being insensitive to the magnetic field and more cost-effective, but with similar timing resolution and comparable radiation length (with respect to 2 cm thick scintillators).

## 5. – Acknowledgements

The use of the instrumented beam line and data acquisition system installed by the ALICE experiment for the tests at the T10 PS beam is acknowledged, as well as the kind support of the CERN EP/AIT group and of André Braem (CERN EP/AT1).

We obviously benefited from many discussions and from the accumulated experience of the ALICE-TOF project.

This work was supported by the portuguese research contract CERN/FAE/1197/98.

## 6. – References